\documentclass[12pt,twoside]{article}

\pdfoutput=1

\usepackage[usenames]{color}
\usepackage{lscape}
\usepackage{amssymb}
\usepackage{rotating}
\usepackage[T1]{fontenc}
\usepackage{graphicx}
\usepackage{fancyhdr}
\usepackage{amsmath}
\usepackage{ae}
\usepackage{aecompl}
\usepackage{hyperref}

\def\nn{\nonumber \\}

\newcommand{\be}{\begin{equation}}
\newcommand{\ee}{\end{equation}}
\newcommand{\ba}{\begin{eqnarray}}
\newcommand{\ea}{\end{eqnarray}}

\newlength{\dinwidth}
\newlength{\dinmargin}
\begin{document}

\thispagestyle{empty}

\begin{flushright}
 EFI-16-24
\end{flushright}

\vspace*{1cm}

\centerline{\Large\bf Enhanced Higgs associated production} 
\centerline{\Large\bf with a top quark pair in the NMSSM with light singlets}

\vspace*{15mm}

\centerline{Marcin Badziak$^{a,b}$ and Carlos E.~M.~Wagner$^{c,d,e}$}
\vspace*{5mm}

\centerline{${}^a$\em Institute of Theoretical Physics,
Faculty of Physics, University of Warsaw} 
\centerline{\em ul.~Pasteura 5, PL--02--093 Warsaw, Poland} 
\centerline{${}^b$\em Berkeley Center for Theoretical Physics, Department of Physics,}
\centerline{\em and Theoretical Physics Group, Lawrence Berkeley National Laboratory,}
\centerline{\em University of California, Berkeley, CA 94720, USA}
\centerline{$^c$\em Enrico Fermi Institute, University of Chicago, Chicago, IL 60637, USA}

\centerline{$^d$\em High Energy Physics Division, Argonne National Laboratory, Argonne, IL 60439, USA}

\centerline{$^e$\em Kavli Institute for Cosmological Physics, University of Chicago, Chicago, IL 60637, USA}

\vskip 1cm

\centerline{\bf Abstract}

Precision measurements of the  125~GeV Higgs resonance recently discovered at the
LHC have determined that its properties are similar to  the ones of the Standard Model (SM) Higgs boson.  However,
the current uncertainties in the determination of the Higgs boson couplings leave room for significant deviations from the
SM expectations.  In fact, if one assumes no  correlation between the top-quark and gluon couplings to the
Higgs, the current global fit to the Higgs data lead to central values of the Higgs couplings to the bottom-quark and
the top-quark that are about 2~$\sigma$ away from the SM predictions. In a previous work, we showed that 
such a scenario could be realized in the Next to Minimal Supersymmetric extension of the SM (NMSSM), for heavy singlets 
and  light MSSM-like Higgs bosons
and scalar top quarks, but for couplings that ruined the perturbative consistency of the theory up to the GUT scale. In this work
we show that a perturbative consistent scenario, for somewhat heavier stops,  may be obtained in the presence of light singlets.
An interesting bonus of this scenario is the possibility of explaining an excess of events observed in CP-even Higgs searches at LEP2.

\vskip 3mm

\newpage

\section{Introduction}

After the discovery of a 125~GeV resonance in July 2012 \cite{Higgsdiscovery}, the LHC collaborations have established that its
properties are close to the ones of the SM Higgs boson, namely a neutral CP-even scalar. The ratio of the observed 
Higgs production cross section to the SM predicted values differs from one by just a few tens of percent in
most final state channels \cite{Higgscomb}. This indicates that the couplings to third generation fermions and vector gauge bosons 
are roughly consistent with their SM predicted values.  The measurement of the Higgs production rate in association
with top quarks in multilepton channels, however, shows central values that are significantly above the SM 
expectations \cite{tthexp}.  Moreover, the search for Higgs bosons produced in association with weak gauge bosons and decaying
into bottom quarks have revealed values that tend to be significantly smaller than the SM predicted values. 
In fact, if one ignores the correlation between the top-quark and gluon couplings to the Higgs, the best fit
to the Higgs couplings leads to central values of the Higgs couplings to bottom-quarks and top-quarks that
are 2~$\sigma$ away from the SM predicted values, as shown by the recent analysis of the Higgs data
by the CMS and ATLAS collaborations \cite{Higgscomb}.   Although these channels are statistically limited, as reflected by
the fact that no Higgs discovery may be established in the $t\bar{t}h$ channel at this point, it is worthwhile
to analyze if such deviations from the SM predictions may be realized in any realistic low energy extension
of the SM.

As has been shown in Ref.~\cite{MBCWtth},  a Higgs with enhanced couplings to the top quark and suppressed couplings
to the bottom quark may be easily obtained in two Higgs doublet models (2HDM), for values of $\tan\beta$, the ratio of
vacuum expectation value, close to one.  The trouble with these models is that the coupling of the Higgs to
gluons is in first approximation proportional to the Higgs coupling to the top quark and therefore one would
expect enhanced gluon fusion production rates too, what is in conflict with experiment.  In addition, the
bottom coupling suppression would lead to a reduction of the width of the Higgs decay into bottom quarks,
and to a subsequent enhancement of the branching ratios  of the Higgs decay into
vector gauge bosons. 

A consistent solution to this problem may be obtained in the presence of additional light color degrees of
freedom, with significant couplings to the Higgs and that may lead to contributions to the gluon coupling
that compensates the top-quark ones.
\footnote{The degeneracy in the gluon fusion production cross-section between the top quark and New Physics contributions can be broken by studying
production of a boosted Higgs with a jet, see e.g.~\cite{Azatov:2013xha,Grojean:2013nya}.}
A particular example is the case of low-energy supersymmetry~\cite{Martin:1997ns},
in which these new colored particles are just given by the stops, the superpartners of the top quark. It
was demonstrated that if the lightest stop mass is of the order of a few hundred GeV and the stop mixing
parameters are large, the Higgs rates may be brought to agreement with the experimentally observed ones~\cite{MBCWtth}.  
This solution cannot be realized in the Minimal Supersymmetric extension of the SM (MSSM) since it is difficult to obtain
the right Higgs mass for such small values of $\tan\beta$ and the stop masses and, in addition, the Higgs mixing effects
in this model lead to an enhancement of the Higgs bottom coupling and a suppression of the top one,
that is the opposite as the tendency indicated by data.  

In Ref.~\cite{MBCWtth} it was shown that this problem may be fixed in the simplest extension of the MSSM, with the addition
of a singlet superfield, namely
the NMSSM~\cite{reviewEllwanger}. In such a case, the same coupling $\lambda$ that enhances the Higgs mass modifies
the CP-even Higgs mixing, leading to couplings of the Higgs consistent with the observed ones. It was shown that, for heavy
singlets, the required values of $\lambda$, the superpotential coupling of the singlet to the Higgs doublet superfields, were 
$\lambda > 0.7$, what leads to the breakdown of the perturbative
consistency of the theory below the GUT scale.  In this article, we shall show that an alternative
solution can be obtained for the case of light singlets. In such a case, the values of $\lambda$ can
be lower than $0.7$ and a perturbative consistent solution can be found.  In addition, although this
is not a requirement of this scenario, we shall demonstrate that values of the masses and
couplings to vector bosons of the (predominantly singlet) lightest CP-even Higgs, consistent
with the ones required to explain an observed excess in the LEP2 Higgs search data \cite{LEP2excess}, can be
obtained in this scenario.

The possibility of a $tth$ coupling enhancement was also studied in the context of models with vector-like top quark partners that mix with the top
quark~\cite{Angelescu:2015kga}. 
The $t\bar{t}h$ production excess is mainly driven by the multi-lepton channel and is one of several excesses in searches in final states consisting 
of bottom quarks and many leptons~\cite{Chatrchyan:2013fea},\cite{Aad:2014pda},\cite{Aad:2015gdg}. 
Such excesses were also explained in models with standard Higgs sector by introducing new scalar or fermion
particles~\cite{Huang:2015fba,Chen:2015jmn,Cheng:2016npb}. 

This article is organized as follow. In section \ref{sec:enhtth} we review the conditions to obtain an enhancement
of the Higgs-top-quark coupling in the presence of light singlets. In section \ref{sec:NMSSM} we present the NMSSM
realization and in section \ref{sec:Higgssignals} we present a numerical analysis of the Higgs production rates in this scenario.
In section \ref{sec:Higgspheno} we analyze the phenomenology of the non-standard Higgs bosons. We reserve section \ref{sec:concl}
for our conclusions.

\section{Enhanced $tth$ coupling with a light singlet}
\label{sec:enhtth}

In order to enhance the associated production of $h$ with a top quark pair (which we shall denote $tth$)
in a model with two Higgs doublets the 125 GeV Higgs eigenstate must have a non-negligible component in the
non-SM-like
doublet. In type-II 2HDM the CP-even Higgs couplings to fermions and gauge bosons (normalized to the SM values)  are determined by 
the values of the CP-even Higgs mixing angle $\alpha$ and $\tan\beta$, the ratio of the two Higgs doublet vacuum expectation values in the
following way :
\begin{align}
\label{eq:ct}
&c_t=\frac{\cos\alpha}{\sin\beta}=\sin\left(\beta-\alpha\right)\left[ 1 + \cot\beta \cot\left(\beta-\alpha\right) \right] \,, \\
\label{eq:cb}
 &c_b=-\frac{\sin\alpha}{\cos\beta}=\sin\left(\beta-\alpha\right) \left[ 1 - \tan\beta \cos\left(\beta-\alpha\right) \right] \,, \\
 \label{eq:cV}
 &c_V=\sin\left(\beta-\alpha\right) \, ,
\end{align}  
where $c_i = g_{hii}/g_{hii}^{\rm SM}$ denotes the ratio of the Higgs coupling to the $i$ particle normalized to its SM value.
Enhanced $tth$ is obtained for low $\tan\beta$ and sizeable $\cot\left(\beta-\alpha\right)>0$. In a pure 2HDM, in which the coupling of the Higgs
bosons to gluons is induced by loops of third generation quarks,  enhanced $tth$ is inevitably
correlated with enhanced Higgs production in the gluon fusion channel, which is phenomenologically unacceptable. One can avoid this correlation if
there exist additional light colored states, such as stops, that give negative contribution to the effective Higgs coupling to gluons~\cite{MBCWtth}.
It was demonstrated in Ref.~\cite{MBCWtth} that in such a case enhancement of the $tth$ signal of the Higgs decaying to gauge bosons can be larger than in the SM by a factor of two without violation of any experimental constraints. Moreover, it was shown in Ref.~\cite{MBCWtth} that this can be also
realized in the NMSSM with heavy singlets, which effectively corresponds to a type-II 2HDM, with relatively light highly-mixed stops.  The stop contributions modify the effective Higgs coupling to gluons and photons in the following way, see e.g.~Refs.\cite{Falkowski,Djouadi}:
\begin{equation}
 \frac{c_g}{c_g^{\rm SM}}=\frac{c_{\gamma}}{c_{\gamma}^{\rm SM}}=c_t + \frac{m_t^2}{4}\left[c_t\left( \frac{1}{m_{\tilde{t}_1}^2} +
\frac{1}{m_{\tilde{t}_2}^2}\right) - \frac{\tilde{X}_t^2}{m_{\tilde{t}_1}^2 m_{\tilde{t}_2}^2} \right] \,,
\label{eq:cg_stopeff}
\end{equation}
where $\tilde{X}_t^2 \equiv X_t  \left( A_t \frac{\cos\alpha}{\sin\beta} + \mu \frac{\sin\alpha}{\sin\beta} \right)$ with the stop mixing parameter
given by $X_t \equiv A_t-\mu/\tan\beta$ (note: in the decoupling limit $\tilde{X}_t^2 = X_t^2$). It should be clear from the above formula that
significant reduction of the Higgs coupling to gluons is possible only for light enough stops. At the ICHEP 2016 conference, the  LHC collaborations presented new constraints on the lightest stop mass, which are quite strong in simplified models in which the mass difference between the stop and
the lightest supersymmetric particle is large~\cite{stopATLAS1,stopATLAS2,stopATLAS3,stopCMS1,stopCMS2}. However, the constraints are still relatively weak for small mass splitting between the stop and the LSP. In order to avoid constraints on light sbottoms it is preferred that the lightest stop is mostly right-handed. The limit is especially weak if the lightest stop decays mainly to charm and neutralino. For such topology a lightest stop as light
as about 260 GeV is allowed if the stop-LSP mass splitting is above about 20 GeV (but not large enough to make the stop decays to charm and
neutralino subdominant)~\cite{stopATLAS_charmneutralino,stopCMS_charmneutralino,stopATLAS8TeV,stopCMS8TeV}. 

Another important requirement to induce a large $tth$ enhancement in the NMSSM with heavy singlets is that the value of 
$\lambda \gtrsim 0.8$~\cite{MBCWtth}. Such large values of $\lambda$ lead to a Landau pole below the Grand Unification (GUT)
scale, $M_{\rm GUT} \simeq 2 \ 10^{16}$~GeV. In the following, we will show that smaller values of $\lambda$ are possible if the singlets are light.

If in addition to the CP-even Higgs bosons proceeding from the  two Higgs doublets also a singlet is light, in order to determine non-SM doublet
component of the 125 GeV Higgs one needs to consider three-by-three Higgs mass matrix (in the Higgs basis):\footnote{The Higgs basis $(\hat{h},
\hat{H}, \hat{s})$ is defined as $\hat{h}=H_d\cos\beta + H_u\sin\beta$, $\hat{H}=H_d\sin\beta -
H_u\cos\beta$ and $\hat{s}=S$. In this basis the $\hat{h}$ field has exactly the same couplings to the gauge bosons and fermions as the SM Higgs field.
The field
$\hat{H}$ is a non-SM-like doublet which does not couple to the gauge bosons and its couplings to the down and up fermions are the SM Higgs ones
rescaled by $\tan\beta$ and $-\cot\beta$,
respectively. The mass eigenstates are denoted as $s$, $h$, $H$, with the understanding that $h$ is the SM-like Higgs. } 
\begin{equation}
 \hat{M}^2=
\left(
\begin{array}{ccc}
  \hat{M}^2_{hh} & \hat{M}^2_{hH} & \hat{M}^2_{hs} \\[4pt]
   \hat{M}^2_{hH} & \hat{M}^2_{HH} & \hat{M}^2_{Hs} \\[4pt]
   \hat{M}^2_{hs} & \hat{M}^2_{Hs} & \hat{M}^2_{ss} \\
\end{array}
\right) \,.
\end{equation}
Since the singlet does not couple to SM particles, the couplings of the 125 GeV are still given by eqs.~\eqref{eq:ct}-\eqref{eq:cV} but the effective
$\cot\left(\beta-\alpha\right)$ depends now also on the mixing of the singlet with the Higgs doublets:
\begin{equation}
\label{eq:cotb_singlet}
 \cot\left(\beta-\alpha\right) =  \frac{  \left(m_h^2 - \hat{M}_{hh}^2\right) \hat{M}_{Hs}^2 + \hat{M}_{hs}^2 \hat{M}_{hH}^2  }{  \left(
\hat{M}_{HH}^2 - m_h^2\right) \hat{M}_{hs}^2 - \hat{M}_{Hs}^2 \hat{M}_{hH}^2 } \,.
\end{equation}
In the above the effective $\sin(\beta-\alpha)$ and $-\cos(\beta-\alpha)$ are defined as the $\hat{h}$ and $\hat{H}$ components of $h$, 
respectively. 
We see that it is possible to generate non-zero $\cot\left(\beta-\alpha\right)$ even for $\hat{M}_{hH}^2=0$. This feature is particularly important
in the context of the NMSSM since for the most interesting values of $\lambda\approx0.65$ (which give a large positive correction to the Higgs mass but do
not generate the Landau pole below the GUT scale) $\hat{M}_{hH}^2\approx0$ which is known as the NMSSM alignment limit~\cite{alignment}. Notice also that
in the limit $\hat{M}_{hH}^2=0$, $tth$ is enhanced for $\hat{M}_{hs}^2 \hat{M}_{Hs}^2 >0$ $(<0)$ if the singlet-like
scalar is lighter (heavier) than 125 GeV. This is because in this limit mixing of the Higgs with lighter (heavier) singlet-like scalar pushes up
(down) the Higgs mass so $\left(m_h^2 - \hat{M}_{hh}^2\right)$ is positive (negative) \cite{Jeong:2012ma,Agashe:2012zq,nmssmmixing}.  It is also
interesting to note that in
principle large $\cot\left(\beta-\alpha\right)$ may be also obtained for very heavy non-SM Higgs doublet if $|\hat{M}_{Hs}^2|$ is large enough.

\section{NMSSM realization}
\label{sec:NMSSM}

Let us now focus on $tth$ enhancement in the general NMSSM for which the MSSM superpotential is supplemented by:
\begin{equation}
\label{W_NMSSM}
 W_{\rm NMSSM}= \lambda SH_uH_d + \xi_F S+\mu'S^2/2+\kappa S^3/3 \,.
\end{equation}
and the soft terms are given by
\begin{align}
-{\cal{L}}_{\rm soft}
\supset
&\,\,
m_{H_u}^2\left|H_u\right|^2+m_{H_d}^2\left|H_d\right|^2+m_{S}^2\left|S\right|^2
\nn
&\label{Lsoft}
+\left(
A_\lambda\lambda H_u H_d S +\frac13A_\kappa\kappa S^3
+m_3^2H_uH_d + \frac12 m_S'^2S^2 + \xi_SS
+{\rm h.c.}\right)\,,
\end{align}
where $S$ is a SM-singlet superfield which scalar component acquires vacuum expectation value $v_s$. 
The first term is the source of the effective higgsino mass parameter, $\mu_{\rm eff}\equiv\lambda v_s$ (we drop the subscript ``eff'' in the rest of
the paper and we set explicitly the  MSSM $\mu$-term to zero by shifting the value of $S$, what amounts to a simple redefinition of some of the 
parameters).

The off-diagonal terms of the tree-level Higgs mass matrix in the Higgs basis in the general NMSSM are given by:
\begin{align}
\label{MhHfull}
& \hat{M}^2_{hH} = \frac{1}{2}(M^2_Z-\lambda^2 v^2)\sin4\beta \,, \\
\label{Mhs}
& \hat{M}^2_{hs} =  \lambda v (2\mu-\Lambda \sin2\beta) \,, \\
\label{MHs}
& \hat{M}^2_{Hs} = \lambda v \Lambda \cos2\beta \,.
\end{align}
where  $\Lambda\equiv A_{\lambda}+\mu'+2\kappa v_s$,  $v\approx174$ GeV, and we have ignored loop-corrections
that  are generically small and have a minor phenomenological impact in the region of parameters relevant for this work~\cite{alignment}.  The diagonal mass in the SM-like Higgs component is given by: 
\begin{equation}
\label{Mhh}
\hat{M}_{hh}^2 = M_Z^2 \cos^2(2\beta) + \lambda^2 v^2 \sin^2(2\beta) + \Delta_{\rm loop} \,,
\end{equation}
where $\Delta_{\rm loop}$ parameterizes loop corrections which are dominated by stops.

The explicit form of the remaining diagonal entries of the Higgs mass matrix, that can be found e.g. in Ref.~\cite{nmssmmixing}, is not relevant for our discussion, since in the 
general NMSSM there is enough
freedom in the parameter space to set them to arbitrary values. 

Eq.~\eqref{Mhh} implies that for small values of $\tan\beta$ the Higgs mass of 125 GeV can be accommodated only for relatively large $\lambda\gtrsim0.5$ unless
stops are heavy. On the other hand, avoiding a Landau pole below the GUT scale sets requires $\lambda\lesssim0.7$ (for $\kappa\ll\lambda$) with
the upper bound getting stronger as $\tan\beta$ decreases. For this range of $\lambda$ the lack of a Landau pole below the GUT scale for the top Yukawa
coupling requires also $\tan\beta\gtrsim1.5$. As a result, in this range of parameters $\hat{M}^2_{hH}\approx0$ so using eqs.~\eqref{Mhs}-\eqref{MHs}
together with eq.~\eqref{eq:cotb_singlet}  one obtains the following approximate formula for the effective $\cot\left(\beta-\alpha\right)$:
\begin{equation}
\label{eq:cotb_singletNMSSM}
 \cot\left(\beta-\alpha\right) \approx  \frac{  \left(m_h^2 - \hat{M}_{hh}^2\right)\lambda v \Lambda \cos2\beta   }{  \left(
\hat{M}_{HH}^2 - m_h^2\right) \hat{M}_{hs}^2  }
\approx
{\rm sgn} \left(\frac{\Lambda \sin2\beta-2\mu}{m_h-m_s} \right)
\bar{g}_s \frac{  \lambda v \Lambda \cos2\beta   }{  
\hat{M}_{HH}^2 - m_h^2 } \,,
\end{equation}
where $\bar{g}_s$ is the $s$ coupling to the $Z$ boson normalized to the corresponding  coupling of the SM Higgs with the same mass. In the
approximation made on the very
right hand side of the above equation we assumed $\bar{g}_s^2\ll 1$ which is justified since the Higgs-singlet mixing must be relatively small to
comply with constraints from the LHC and LEP.
Smallness of the Higgs-singlet
mixing requires $|\hat{M}^2_{hs}|\ll m_h^2 + m_s^2$ which is fulfilled when
\begin{equation}
\label{eq:hsalignment}
 \Lambda \approx \frac{2\mu}{\sin2\beta} \,,
\end{equation}
and implies in particular that $\mu\Lambda$ must be positive. We should stress, however, that the above condition should not be
satisfied exactly
because otherwise $\cot\left(\beta-\alpha\right)$ would vanish. Using the above condition together with eq.~\eqref{eq:cotb_singletNMSSM} one obtains
an approximate formula for the effective $\cot\left(\beta-\alpha\right)$ as a function of $\mu$, namely
\begin{equation}
\label{eq:cotb_singletNMSSMmu}
 \cot\left(\beta-\alpha\right) 
\approx
{\rm sgn} \left(\frac{\Lambda \sin2\beta-2\mu}{m_h-m_s} \right)
\bar{g}_s \frac{  \lambda v \mu \cot2\beta   }{  
\hat{M}_{HH}^2 - m_h^2 } \,.
\end{equation}
In order to enhance $tth$ signal one has to also guarantee that $\cot\left(\beta-\alpha\right)>0$ which happens if the following condition is
fulfilled:
\begin{equation}
 |\Lambda| > \frac{2|\mu|}{\sin2\beta} \;\;\;\;\;\;\;\;\;\;\;\;\; \, \left(  |\Lambda| < \frac{2|\mu|}{\sin2\beta}  \right)
\end{equation}
for $m_s<m_h$ ($m_s > m_h$).

Eq.~\eqref{eq:cotb_singletNMSSMmu} confirms the  intuitive expectation that large $tth$ enhancement prefers relatively light MSSM-like Higgs bosons.
However, it also shows that large $|\mu|$ is preferred and that $H$ does not have to be very light  if $|\mu|$ is large
enough. The importance of $|\mu|$ can be seen from table~\ref{tab:bench1} where several benchmark points with large $tth$ enhancement are presented.  We shall define the theoretically predicted signal strengths modifiers as:
\begin{equation}
 R_i^j\equiv\frac{\sigma^j\times {\rm BR}(h \to i)}{\sigma^{j{\rm SM}}\times {\rm BR}^{\rm SM}(h \to i)}  , \,.
\end{equation}
and we shall distinguish the theoretical predictions for the signal strengths from the corresponding LHC measurements, that we define in the conventional way as $\mu_i^j$. 
Comparing points P3 and P4 we see  that similar $tth$ enhancement is
possible for $M_A=300$ GeV and $\mu=500$ GeV as for $M_A=400$ GeV and $\mu=650$ GeV with comparable Higgs-singlet mixing.
Eq.~\eqref{eq:cotb_singletNMSSMmu} also implies that the effective $|\cot\left(\beta-\alpha\right)|$ increases with $\tan\beta$. This is the reason why
points P1 and P3 have similar values of $R^{\rm tth}_{VV}$ in spite of the fact that P3 features smaller Higgs-singlet mixing and the value of
$\mu$ while 
the value of $M_A$
is the same. 

\begin{table}
\centering
\begin{tabular}{c|cccc}
 & {\rm P1} & {\rm P2} & {\rm P3} & {\rm P4}   \\
\hline
$\lambda$  & 0.5 & 0.52 & 0.5 & 0.52 \\
$\tan \beta$  & 1.6 & 1.6 & 2 & 2 \\
\hline
$m_{Q_3}$  & 800 & 800 & 800 & 800\\
$m_{U_3}$  & 240 & 250 & 235 & 200\\
$A_t$  & -1500 & -1500 & -1550 & -1480\\
\hline
$\mu$          & 550 & 650 & 500 & 650 \\
$\mu'$         & 310 & 350 & 235 & 250 \\
$M_{A}$        & 300 & 300 & 300 & 400 \\
$M_{P}$        & 244 & 308 & 297 & 305 \\
$A_{\lambda}$  & 810 & 963 & 908 & 1305 \\
\hline
$m_{s}$                 & 98 & 98 &  88 &  84 \\
$m_{h}$                 & 124.9 & 125.6 & 124.9 & 126.3 \\
$m_{H}$                 & 303 & 336 & 355 & 414 \\
$m_{H^\pm}$             & 217 & 199 & 224 & 317  \\
$m_{a}$                 & 108 & 94 & 105 & 87 \\
$m_{A}$                 & 311 & 358 & 359 & 434 \\
$m_{\tilde{\chi}_1^0}$  & 235 & 235 & 235 & 235  \\
$m_{\tilde{t}_1}$       & 272 & 277 & 276 & 275  \\
$m_{\tilde{t}_2}$       & 946 & 951 & 945 & 948 \\
\hline
$R^{\rm tth}_{VV}$               & 1.60 & 1.62 & 1.60 & 1.60  \\
$R^{\rm tth}_{\gamma\gamma}$     & 1.83 & 1.85 & 1.80 & 1.78   \\
$R^{\rm gg}_{VV}$                & 1.02 & 1.00 & 1.04 & 1.04  \\
$R^{\rm gg}_{\gamma\gamma}$      & 1.16 & 1.15 & 1.16 & 1.16  \\
$R^{\rm VBF/VH}_{VV}$            & 1.32 & 1.34 & 1.41 & 1.39  \\
$R^{\rm VBF/VH}_{\gamma\gamma}$  & 1.51 & 1.54& 1.58 & 1.55 \\
$R^{\rm VBF/VH}_{\tau\tau}$      & 0.72 & 0.76 & 0.77 & 0.73  \\
\hline
$\xi_{b\bar{b}}^{\rm LEP}$       & 0.10 & 0.06 & 0.04 & 0.06 \\
$\bar{g}_s$                      & 0.30 & 0.24 & 0.20 & 0.24 \\
\hline
\end{tabular}
\caption{List of benchmark points obtained with {\tt NMSSMTools 5.0.0} \cite{NMSSMTools}. All masses are in GeV. All points satisfy all experimental
constraints from
the Higgs signal strength
measurements, as well as from direct searches for Higgs bosons, checked with {\tt HiggsBounds 4.3.1}
\cite{HiggsBounds}, and stops. The gluino and the remaining soft sfermion masses are set to 2 TeV,
$M_2=1$ TeV,  $M_1=-235$ GeV. All the remaining $A$-terms are set to $1.5$ TeV, while $\kappa=A_\kappa=m_3^2=m_S'^2=0$. The remaining parameters are
calculated with {\tt NMSSMTools} using EWSB conditions and the values of $\mu$, $M_{A}$ (diagonal mass of MSSM-like pseudoscalar) and $M_P$ (diagonal
mass of singlet-like pseudoscalar). }.
\label{tab:bench1}
\end{table}

Another interesting feature of this scenario is that light singlet-like scalar can explain the LEP2 excess \cite{LEP2excess}. Indeed, the 
LEP2 experiments observed an excess of $b\bar{b}$ events, produced in association with a $Z$ gauge boson,  with an invariant mass in the vicinity of 98~GeV and a signal strength of about one tenth of the one of the SM Higgs 
with the same mass. Phenomenological aspects of the NMSSM with singlet-like scalar explaining the LEP2 excess were studied e.g.~in
Refs.~\cite{BEGJKS,Bhattacherjee:2013vga,Ellwanger:2015uaz} while in Ref.~\cite{nmgmsb} it was shown that this excess can be explained in a UV
complete NMSSM model based on gauge mediated SUSY breaking~\cite{DGS}.
However, none of those works links the LEP2 excess to the $tth$ enhancement. In table~\ref{tab:bench1} we give a value for the prediction of this
signal strength:
\begin{equation}
 \xi_{b\bar{b}}^{\rm LEP}\equiv \bar{g}_s^2 \times \frac{{\rm BR}(s \to b\bar{b})}{{\rm BR}(h^{\rm SM} \to b\bar{b})} \,,
\end{equation}
 Note that the SM normalized $sb\bar{b}$
coupling in the present scenario is enhanced with respect to the corresponding $sZZ$ coupling so $\xi_{b\bar{b}}^{\rm LEP}>\bar{g}_s^2$.
Wee see that point P1 fits very well the LEP2 excess since it features $m_s \approx 98$ GeV and $\xi_{b\bar{b}}^{\rm LEP}\approx0.1$. Point P2 also
has $m_s \approx 98$ GeV but smaller Higgs-singlet mixing, hence also $\xi_{b\bar{b}}^{\rm LEP}$, than P1 so in order to have $tth$ enhancement of
similar size $|\mu|$ is larger in P2 than in P1.

Even though it is an interesting possibility that this scenario can simultaneously explain the $tth$ enhancement and the LEP2 excess, we should
emphasize that our scenario does not require to have the singlet-like scalar mass to be close to the one necessary to explain the LEP excess. 
It is the size of the Higgs-singlet mixing rather than $m_s$ which controls the magnitude of the $tth$ enhancement as can be seen from benchmarks P3
and P4 that feature
$m_s$ far away from the one consistent with the  LEP excess. 
It is noteworthy that $m_s$ can be as small as about 85 GeV (or even smaller if $|\mu|$ is larger than in benchmark P4) without inducing a conflict with
stringent LEP constraints. 

\section{Higgs signal rates}
\label{sec:Higgssignals}

All benchmark points presented in table~\ref{tab:bench1} are compatible with the combination of the run-I Higgs signal measurements~\cite{Higgscomb}
at least at the $2\sigma$ level.
Nevertheless, there are some potential tensions of this scenario with recent run II data that will be probed with future LHC measurements. Let us now discuss these deviations and
how they depend on the model parameters in some more detail. Let us start with the $\gamma\gamma$ decay channel in the gluon fusion production mode.
All benchmarks were chosen to have $R^{\rm gg}_{\gamma\gamma}\approx1.15$ by appropriate adjustment of the stop sector parameters. This is very close
to the central value of the ATLAS and CMS
combination of the Run-I data which yields $\mu^{\rm
gg}_{\gamma\gamma}=1.10^{+0.23}_{-0.22}$ \cite{Higgscomb}. Preliminary results of Run II indicate that both experiments observed some suppression of
this signal strength: $\mu^{\rm
gg}_{\gamma\gamma}=0.59^{+0.29}_{-0.28}$ for ATLAS \cite{ATLAS_hgammagamma_ICHEP} and $\mu^{\rm
gg}_{\gamma\gamma}=0.77^{+0.25}_{-0.23}$ for CMS \cite{CMS_hgammagamma_ICHEP} so the $2\sigma$ upper bound from ATLAS Run II on $R^{\rm
gg}_{\gamma\gamma}$ is about 1.17. By
the end of this year the LHC is expected to deliver a few times more data than analysed for ICHEP2016 so the discrepancy between the Run-I and Run-II
results
should be clarified relatively soon.

While the mechanism that we propose may generate some tension with the Run-II measurement of $\mu^{\rm gg}_{\gamma\gamma}$ the fit to the Run-II
results for $h\to\gamma\gamma$ in the VBF production mode is significantly improved. Indeed, the ATLAS Run-II result is $\mu^{\rm
VBF}_{\gamma\gamma}=2.24^{+0.80}_{-0.71}$ which is almost $2\sigma$ above the SM prediction. It can be seen from the benchmark table that in the
present
scenario it is generically brought to $1\sigma$ agreement with the ATLAS result. Moreover, the prediction is very close to the central value of the
Run-II CMS result of $\mu^{\rm VBF}_{\gamma\gamma}=1.61^{+0.9}_{-0.8}$.

As already emphasized, the present scenario features also suppressed Higgs coupling to down-type fermions. This is in very good agreement with very
weak signal of the Higgs decaying into $b\bar{b}$ observed in both Run I and Run II of the LHC. On the other hand, the Higgs decays to $\tau\tau$
in the VBF production mode (gluon fusion is far less sensitive in this decay channel) has been observed to be relatively close to the SM prediction.
Nevertheless, due to the large uncertainties, values of $R^{\rm VBF/VH}_{\tau\tau}$ as small as about 0.4 (0.8) are consistent with current data at $2\sigma$ ($1\sigma$) level \cite{Higgscomb}. We
can see from table \ref{tab:bench1} that $R^{\rm VBF/VH}_{\tau\tau}$ is in the range between 0.7 and 0.8 for $R^{\rm tth}_{VV}$ of about 1.6. Larger
values of $R^{\rm
VBF/VH}_{\tau\tau}$ keeping the same value of $R^{\rm tth}_{VV}$  are obtained for smaller Higgs-singlet mixing which is evident from comparison
of benchmark P1 with the other ones. Needless to say that $R^{\rm VBF/VH}_{\tau\tau}$ also deviates less from the SM prediction when $R^{\rm
tth}_{VV}$ is smaller. It is also interesting to note that  a given value of $R^{\rm tth}_{VV}$ fitting the LEP2 excess fixes $R^{\rm
VBF/VH}_{\tau\tau}$ (up to small variations from $\tan\beta$ dependence),  e.g.  $R^{\rm tth}_{VV}\approx1.6$ implies $R^{\rm
VBF/VH}_{\tau\tau}\approx0.7$ as for benchmark P1. Such indirect cross-check is particularly important since in this scenario the CP-even singlet $s$ decays are predominantly into $b\bar{b}$ final states.

Let us also note that the larger $\tan\beta$ is the more enhancement of $R^{\rm tth}_{VV}$ comes from suppressed BR$(h\to b\bar{b})$ which results in
closer values for signal rates in $tth$ and VBF/VH production modes.  Therefore,  if the excess in the $tth$ channel persists, information about
$\tan\beta$ may be also extracted from
future Higgs precision measurements. 

We should also comment that all benchmarks presented in table~\ref{tab:bench1} feature rather large values of $|X_t|$ with $|X_t|/m_{\tilde{t}_2}$
around two. Such values are needed to keep the gluon fusion signal rates close to SM predictions when the $tth$ signals are enhanced. It is known from
MSSM studies that too large values of $|A_t|$ may lead to color and/or charge breaking global minima~\cite{CCB_Casas}. Indeed, all the benchmarks in
table~\ref{tab:bench1} feature unphysical global minima. Whether these benchmarks are phenomenologically viable depends on the lifetime of the
metastable EW vacuum. It was recently shown that
constraints on $|A_t|$ from stability of the EW vacuum are overestimated by analytical formulae, presented e.g. in Ref.~\cite{CCB_Casas}, if one
admits as a viable solution
sufficiently long-lived metastable EW vacua~\cite{Vacuum_stop1,Vacuum_stop2,Vacuum_Staub}. It was also emphasized in Ref.~\cite{Vacuum_Staub} that it
is essentially impossible to have a simple universal phenomenological formula that may reliable estimate whether a given point in the MSSM parameter
space leads to destabilization of the EW vacuum and a dedicated study is necessary. Moreover, the MSSM analyses considered only moderate and large
$\tan\beta$, which is not our case. Last but not least, the NMSSM scalar potential has a richer structure than the MSSM one. Studies of color and charge
breaking vacua in the scale-invariant NMSSM have been performed in the past but they either consider NMSSM specific directions~\cite{CCB_NMSSM} in 
field space or give a simple generalizations of the MSSM analytical formulae~\cite{CCB_Ellwanger} which do not account for the fact that EW vacuum may
be metastable and long-lived.  
It is beyond the scope of the present paper to
investigate the conditions for metastability of the EW vacuum in the general NMSSM and we leave a dedicated study of this issue for future work. 
Nevertheless, we expect that for a given value of the lightest stop mass there exist an upper bound on $|X_t|/m_{\tilde{t}_2}$ which could be
translated to an upper bound on possible $tth$ enhancement 
after taking into account the Higgs measurements in the gluon fusion production modes.  
~\\

\section{Non-SM-like Higgs phenomenology}
\label{sec:Higgspheno}

The phenomenology of non-SM-like Higgs bosons is significantly different from that in the case of heavy singlet discussed in Ref.~\cite{MBCWtth}.
Both, light
and heavy singlet scenarios feature heavy doublet-like Higgs bosons in the range of several hundred GeV but their dominant decay channels are totally
different. In the heavy singlet case $H$ and $A$ have large branching ratio for decay into a pair of lightest neutralinos and $H\to hh$ decay is also
very frequent if kinematically accessible. In contrast, in the light singlet scenario $H/A$ decay products in dominant channels involve light
singlet-like (pseudo)scalars, as can be seen from table~\ref{tab:bench2}. The main decay mode for $A$ is $as$ and $Zs$. BR$(A\to H^\pm W^\mp)$ may
also be large if this decay channel is kinematically open.  The main decay channels for $H$ involve  $aa$, $aZ$ and sometimes $ss$ and $H^\pm W^\mp$.
 The only existing analysis
that may probe the sector of heavy neutral
Higgses is the CMS search for $A\to Zs$ or $H\to Za$ with $Z$ decaying leptonically and the lighter pseudo(scalar) decaying to
$b\bar{b}$. Branching
ratios of $A\to Zs$ and $H\to Za$ vary between 10 and 40~\% ~\cite{CMS_AtoZs}. The upper limits on the 13 TeV cross-section for $llb\bar{b}$
production in this
topology are between 0.1 and 1 pb (depending on the masses of the heavy and light Higgs bosons) with 2.3 fb$^{-1}$ of 13 TeV  CMS data. This is at
least
factor of few above the prediction for this cross-section for the benchmark points presented in table~\ref{tab:bench2}.  Therefore,  with ${\mathcal
O}(50)$ 
fb$^{-1}$ it may be possible to probe some of the parameter space that predicts $tth$ enhancement but for a generic point in parameter space this
would require much more data. 
Searches for double Higgs production in $b\bar{b}b\bar{b}$ final state may be more powerful to test this model provided that the experimental
collaborations will relax the assumption that the products of heavy resonance decay have mass of 125 GeV as it is assumed in existing analyses
\cite{Hhh_bbbb}.
It is also noteworthy that $H \to hh$ decays are negligible in the light singlet scenario which is due to the fact that the correct Higgs mass requires
$\lambda\approx0.5$ for which
$Hhh$ coupling is strongly suppressed due to approximate alignment in the $\hat{h}-\hat{H}$ sector \cite{alignment}. 

\begin{table}
\centering
\begin{tabular}{c|cccc}
 & {\rm P5} & {\rm P6}    \\
\hline
$\lambda$  & 0.48 &  0.5 \\
$\tan \beta$  & 1.6  & 2 \\
$\kappa$ & 0.48 & 0.48 \\
\hline
$m_{Q_3}$  & 800 & 800 & \\
$m_{U_3}$  & 195 & 190 \\
$A_t$  & -1400 & -1450 \\
\hline
$\mu$          & 700 & 700  \\
$A_{\lambda}$  & -15 & 223  \\
$A_{\kappa}$  & -3678 & -3431  \\
$\xi_S$   & -$10^9$ & -$10^9$  \\
$m_3^2$   & -$4.9\cdot10^5$ & -$6.1\cdot10^5$  \\
$m_S'^2$   & 0 & -$3.64\cdot10^6$  \\
\hline
$m_{s}$                 & 98 & 90  \\
$m_{h}$                 & 125.1 & 125.5  \\
$m_{H}$                 & 325 & 405 \\
$m_{H^\pm}$             & 246 & 271  \\
$m_{a}$                 & 2834 & 107  \\
$m_{A}$                 & 248 & 442  \\
$m_{\tilde{\chi}_1^0}$  & 235 & 234   \\
$m_{\tilde{t}_1}$       & 271 & 277  \\
$m_{\tilde{t}_2}$       & 950 & 949  \\
\hline
$R^{\rm tth}_{VV}$               & 1.59 & 1.59   \\
$R^{\rm tth}_{\gamma\gamma}$     & 1.81 & 1.76   \\
$R^{\rm gg}_{VV}$                & 1.03 & 1.05   \\
$R^{\rm gg}_{\gamma\gamma}$      & 1.17 & 1.16   \\
$R^{\rm VBF/VH}_{VV}$            & 1.31 & 1.40  \\
$R^{\rm VBF/VH}_{\gamma\gamma}$  & 1.49 & 1.56 \\
$R^{\rm VBF/VH}_{\tau\tau}$      & 0.71 & 0.77   \\
\hline
$\xi_{b\bar{b}}^{\rm LEP}$       & 0.11 & 0.04  \\
$\bar{g}_s$                      & 0.32 & 0.20  \\
\hline
\end{tabular}
\caption{List of benchmark points that avoid unphysical global minimum. Unspecified MSSM-like parameters are the same as in table~\ref{tab:bench1}
while $\mu'=\xi_F=0$. All dimensionful parameters are in GeV except for $\xi_S$ in GeV$^3$ and $m_3^2$ and $m_S'^2$ in GeV$^2$ }.
\label{tab:bench3}
\end{table}

\begin{table}
\centering
\begin{tabular}{c|cccccc}
& {\rm P1} & {\rm P2} & {\rm P3} & {\rm P4}  & {\rm P5} & {\rm P6}  \\
\hline
BR$(H\to t\bar{t})$     & 0    & 0     & 0.01 & 0.07 & 0 & 0.17\\
BR$(H\to ss)$           & 0.32 & <0.01 & 0.05 & 0.28 & 0.72 & 0.01\\
BR$(H\to aa)$           & 0.31 & 0.66  & 0.39 & 0.12 & 0 & 0.81\\
BR$(H\to aZ)$           & 0.22 & 0.19  & 0.35 & 0.38 & 0 & 0.12\\
BR$(H\to hs)$           & 0.11 & <0.01 & <0.01 & 0.11 & 0.24 & <0.01 \\
BR$(H\to H^\pm W^\mp)$  & 0.01 & 0.13  & 0.19 & 0.03 & 0 & 0.05\\
\hline
BR$(A\to t\bar{t})$         & 0 & 0.14 & 0.13 & 0.18 & 0 & 0.06\\
BR$(A\to as)$            & 0.66 & 0.39 & 0.32 &  0.35 & 0 & 0.61 \\
BR$(A\to Zs)$            & 0.22 & 0.23 & 0.33 &  0.32 & 0.95 & 0.14\\
BR$(A\to ah)$            & 0.08 & <0.01 & <0.01 &  0.05 & 0 & 0.07\\
BR$(A\to H^\pm W^\mp)$   & 0.04 & 0.24 & 0.20 &  0.08 & 0 & 0.12\\
\hline
BR$(H^+\to t\bar{b} ) $  & 0.58 & 0.43 & 0.34 & 0.29 & 0.70 & 0.30 \\
BR$(H^+\to W^+ a )$      & 0.19 & 0.34 & 0.27 & 0.38 & 0 & 0.34 \\
BR$(H^+\to W^+ s )$      & 0.23 & 0.23 & 0.39 & 0.32 & 0.29 & 0.36 \\
\hline
$\sigma(ggH)$ [pb]       & 2.4 & 1.6 & 1.2 & 1.4 & 2.7 & 1.2 \\
$\sigma(ggA)$ [pb]       & 4.9 & 5.2 & 4.0 & 2.2 & 11.9 & 1.3 \\
$\sigma(gga)$ [pb]       & 24.3 & 40.4 & 18.7 & 19.2 & $10^{-9}$ & 17.7 \\
$\sigma(ggs)$ [pb]       & 4.6 & 10.6 & 10.3 & 7.1 & 3.9 & 9.8 \\
\hline
$\Omega_{\rm th} h^2$    & 0.08 & 0.09 & 0.07 & 0.06 & 0.08 & 0.07\\
$\sigma_{\rm SI}^{\rm p}$ [pb]    & $3.8\cdot 10^{-10}$ & $3\cdot 10^{-10}$ & $8.4\cdot 10^{-10}$ & $4.3\cdot 10^{-10}$ & $2.5\cdot 10^{-10}$
& $3.8\cdot 10^{-10}$ \\
\end{tabular}
\caption{Branching ratios and gluon-fusion production cross-sections (calculated with {\tt SuShi 1.6.0} \cite{SusHi})  for
non-SM-like Higgs bosons for benchmark points presented in tables~\ref{tab:bench1} and \ref{tab:bench3}. Thermal relic abundance $\Omega_{\rm th} h^2$
of the LSP and
SI LSP-proton scattering cross-section $\sigma_{\rm SI}^{\rm p}$ are also
given.}.
\label{tab:bench2}
\end{table}

The charged Higgs is also relatively light. It is typically not far above $H^{\pm}\to tb$ threshold while the ATLAS Run-II results for charged Higgs
search in $t\bar{b}$ decay mode 
are available only for $m_{H^\pm}>300$ GeV with the strongest upper bound on $\tan\beta$ in type-II 2HDM with BR$(H^{\pm}\to
tb)=100 \%$ of about 1.7
for $m_{H^\pm}=300$ GeV \cite{chargedHiggs_Atlas13}. While the charged Higgs production cross-section in our model is expected to be similar as in the
type-II 2HDM,
BR$(H^{\pm}\to
tb)$ is much less than 100~\% since charged Higgs often decays to $W^+ a$ and $W^+ s$ which weakens the bound on $\tan\beta$ significantly.  
This is the reason why benchmark P4 which features $m_{H^\pm}>300$ GeV comfortably satisfies current constraints. Nevertheless, future 
searches for charged Higgs in $t\bar{b}$ decay mode may probe important part of the model parameter space, especially if they are extended to masses
below 300 GeV. While there are difficulties in reliable computation of the charged Higgs production cross-section for masses close to the top quark
mass, some progress has been made on this front recently \cite{Degrande:2016hyf}, so comparing the experimental results with theoretical predictions may
be easier now. Moreover, it was argued in Ref.~\cite{chargedHiggs_NMSSM_Wa} that the NMSSM charged Higgs decaying to $W^+ a$ could be efficiently
probed in searches for events with missing transverse momentum, b-jets, leptons and/or taus.

As already emphasized, the crucial feature that allows for significant $tth$ enhancement is large mixing between singlet scalar and MSSM-like
doublet scalar. This generically implies also large mixing in the pseudoscalar sector. Indeed, for benchmarks presented in tables
\ref{tab:bench1}-\ref{tab:bench2} this mixing is close to maximal. One consequence of this is large gluon-fusion production cross-section for lighter
pseudoscalar, as seen from table~\ref{tab:bench2}. Direct production of $a$ may be discovered at the LHC in the $\tau\tau$ decay channel which
contributes about 10~\% to total decay width of $a$ (the remaining 90~\% of $a$ decays are to $b\bar{b}$ but this decay channel is 
extremely difficult to observe in the
gluon-fusion production mode at the LHC). On the other hand, CMS $\tau\tau$ search extends down to masses of about 90 GeV, so it covers an interesting part of parameter space of
the model including benchmarks P1-P3. The current upper limit on $\sigma(gga)\times{\rm BR}(a\to\tau\tau)$ from the 13 TeV CMS search with 2.3
fb$^{-1}$
of data \cite{CMStautau_13TeV} is about 40 pb for the mass close to 100 GeV
with the limit improving rather quickly for
masses above about 120 GeV.
Therefore, the limit is an order of magnitude above the benchmark predictions so while some of the parameter space may be probed with 300 fb$^{-1}$,
as exemplified by P2, vast of the parameter space will require high-luminosity LHC to be covered.

Weaker constraints for a ditau resonance with mass about 100 GeV are
also the reason why it is hard to probe the production of the  98 GeV singlet scalar explaining the LEP excess. 
For benchmark P1 which fits the LEP excess the cross-section for $\tau\tau$ from decay of $s$ produced in gluon fusion  is about 100 times smaller
than the CMS upper limit so it might not be probed even
at high-luminosity LHC. It is interesting to note that the benchmark P2 has larger gluon-fusion production cross-section of $s$ than P1 by more than a
factor of two even though $s$ has much smaller $\hat{h}$ component than in P1. This is mainly because $\hat{h}$ and $\hat{H}$ components of $s$ give
opposite sign contributions to $s$ coupling to top quarks,  which is the main source of effective coupling to gluons. It turns out that these
contributions have similar magnitude for values of $\bar{g}_s$ that can explain the LEP excess which results in a small $s$ coupling to top quarks, and
hence also to gluons. For smaller values of $\bar{g}_s$ this cancellation is less efficient and a larger $\hat{H}$-$\hat{s}$ mixing is needed to keep
the
same $tth$ enhancement. In consequence, larger cross-sections for $s$ production via gluon fusion are predicted. 

Since scalars, unlike pseudoscalars, couple to weak gauge bosons at tree-level, $s$ can be also produced in VBF or VH modes. These modes have smaller production cross sections but also
suffer from less background and for 125 GeV scalars they turn out to be much more sensitive to the gluon-fusion mode both in $b\bar{b}$ and
$\tau\tau$ channels.
Unfortunately, the LHC collaborations have not looked at these production modes for masses below 100 GeV so currently they cannot test the LEP2 excess.
 We strongly encourage them to extend their analyses to lower masses.

Let us also comment on the fact that in this scenario the LSP may be a good dark matter candidate. In the presented benchmarks the LSP is mainly a 
Bino, but its thermal relic abundance $\Omega_{\rm th} h^2$ (computed with {\tt microOMEGAs} \cite{micromegas}) is in agreement with the upper bound
on $\Omega h^2$ from
Planck measurements \cite{Planck}. This is because small mass
splitting between the LSP and the lightest stop required to satisfy the LHC constraints results in efficient stop co-annihilations. Non-negligible LSP
annihilation
into final states involving light (pseudo)scalars is another reason for $\Omega_{\rm th} h^2$ much smaller than generically predicted for
Bino-dominated LSP. In fact, for the benchmarks $\Omega_{\rm th} h^2$ ranges between 0.06 to 0.1 so is somewhat below the central value of
0.12 measured by Planck \cite{Planck}. Larger values of $\Omega_{\rm th} h^2$ may be obtained by increasing $\mu$ which leads to reduced Higgsino
component of the LSP, hence smaller annihilation cross-section.
The LSP scattering cross-section on nucleons for all the benchmarks is below the LUX limits \cite{LUX} but within the future reach of Xenon1T for
direct detection via spin-independent (SI) LSP-nucleon interactions \cite{Xenon1T}. The compatibility with the LUX constraints is the reason for
choosing negative value of $M_1$ in the benchmarks because $\mu M_1<0$ allows for some cancellations in the Higgs coupling to LSP which results in
smaller SI scattering cross-section \cite{bsMSSM}.  Additional suppression of the SI scattering cross-section may originate from destructive
interference between the contributions to the SI scattering amplitude from the Higgs and the light singlet-scalar exchange which are of comparable
size if Higgs-singlet mixing is non-negligible as in the present scenario \cite{bsNMSSM}.  For somewhat larger $\tan\beta$ SI scattering
cross-section could be also suppressed
by effects of destructive interference with the amplitude generated by MSSM-like Higgs exchange~\cite{bsMA}. 
Due to small fraction of higgsino component of the LSP, hence small LSP coupling to Z boson, the spin-dependent (SD) scattering cross-section on
neutrons is in the range between $10^{-7}$ and $10^{-6}$ pb so about three order of magnitude below a recent LUX limit \cite{LUXSD}.  Limits from
the SD scattering cross-section on protons are even weaker.

Since this scenario requires
 light highly-mixed stops and many additional light Higgs bosons there are non-negligible contributions to $B$-physics observables. We computed these
observables with {\tt NMSSMTools} that follows the calculation of Ref.~\cite{NMSSMDomingo}, which assumes Minimal Flavor Violation (MFV). 
For all benchmarks ${\rm BR}(b \to s \gamma)$,  ${\rm
BR}(B_s \to  \mu^+ \mu^-)$ and ${\rm BR}(B_d \to  \mu^+ \mu^-)$ are below the SM prediction while ${\rm
BR}(B \to X_s \mu^+ \mu^-)$ (in the low dimuon invariant mass region) is above the SM prediction.  The ${\rm BR}(b \to s \gamma)$ is in agreement
with the experimental central value \cite{bsg_exp} at the 
$2\sigma$ level for all points apart from P2, while a similar discrepancy with the experiment for ${\rm
BR}(B \to X_s \mu^+ \mu^-)$ is observed only for P1 and P2 \cite{BXsmumu}. This proves that it is possible to find points featuring $tth$
enhancement consistent with these observables, even under the Minimal Flavor Violation (MFV) assumption.  It should be also emphasized 
that the predictions for $B$-physics
observables are sensitive to sources of flavor violation beyond MFV, for example to flavor structure of the
down squark parameters via loops with gluinos \cite{Gabbiani:1996hi}, that have a minor effect in Higgs physics, that is the subject of this work. 
The only measurement that is in more than $2\sigma$ tension with all the benchmarks in the MFV scenario is the combined CMS
and LHCb analysis of $B_s \to \mu^+ \mu^-$ and $B_d \to  \mu^+ \mu^-$ decays \cite{Bmumu_CMSLHCb}. However, this is due to the fact that the SM
prediction is already $2\sigma$ away from the experimental central value. Moreover,  in
Ref.~\cite{Bmumu_Rosiek} it was specifically demonstrated that ${\rm
BR}(B_s \to  \mu^+ \mu^-)$ and ${\rm BR}(B_d \to  \mu^+ \mu^-)$ are very sensitive to non-MFV parameters, especially for low $\tan\beta$.

We should also warn the reader that benchmarks P1-P4 presented in Table~\ref{tab:bench1} possess unphysical global minimum characterised by vanishing
$\langle S\rangle$ and $\langle H_u\rangle$ (or $\langle H_d\rangle$). Such minima have a depth $-2m_{H_d}^4/g^2$ (or $-2m_{H_u}^4/g^2$) and were
analysed  in the $\mathbb{Z}_3$-invariant NMSSM in Refs.\cite{CCB_NMSSM,Ellwanger:1996gw}. It is beyond the scope of the present paper to calculate
the lifetime of the metastable EW vacuum for these benchmarks. However, in the general NMSSM there is enough freedom to make the EW vacuum deeper than
the above mentioned unphysical minima while preserving characteristic features of benchmarks presented in table~\ref{tab:bench1}. For example,
large negative $\xi_S$ can make the EW vacuum deeper than the unphysical ones. Such values of $\xi_S$ give positive contribution to the CP-even
singlet mass but negative value of $\kappa A_\kappa$ can keep the singlet light. In table~\ref{tab:bench3} we present benchmarks with the EW vacuum
deeper than the unphysical ones. P5 has the same qualitative features as P1 except for heavy CP-odd singlet which results from large
negative $\xi_S$. In consequence, only the CP-even singlet is present in decays of MSSM-like Higgs bosons, as seen from table~\ref{tab:bench2}.
Nevertheless, the
CP-odd singlet can be light also in this case for appropriately chosen positive value of $m_S'^2$, as exemplified by benchmark P6.

\section{Conclusions}
\label{sec:concl}

The analysis of the Higgs precision measurement data at run I of the LHC have revealed large uncertainties in the coupling of the
recently observed Higgs boson to the third generation quarks. In particular, the best fit value of the Higgs coupling to
top quarks is more than 2~$\sigma$ above the SM value. Similarly, the best fit of the Higgs coupling to bottom quarks is 2~$\sigma$ 
below the SM value. 

In this article we have investigated the possibility of obtaining such modifications of the third generation couplings within
supersymmetric extensions of the Standard Model. In particular we have shown that they may be obtained in the 
NMSSM for values of the singlet state masses below  125~GeV and values of the heavy CP-even Higgs mass of the
order of a few hundred GeV. In addition, in order to get consistency with the observed gluon fusion rates,  light stops 
and relatively large values of the stop mixing parameters are also required.  We have discussed the phenomenological
implications of this scenario, showing, for instance that it may be also compatible with the observed LEP2 excess in the search
for associated production of Higgs bosons with neutral gauge bosons, $e^+e^- \to h Z$. 

This scenario predicts not only deviations of the Higgs rates with respect to the SM values, but also a rich pattern of non-standard Higgs decays,
many of which are not covered by existing LHC searches. 
It will be also tested in the search for light stops, and therefore may be probed in multiple
channels at the LHC in the near future.

\section*{Acknowledgments}
This work has been partially supported by National Science Centre under research grant DEC-2014/15/B/ST2/02157, by
the Office of High Energy Physics of the U.S. Department of Energy
under Contract DE-AC02-05CH11231, and by the National Science Foundation
under grant PHY-1316783. MB acknowledges support from the
Polish 
Ministry of Science and Higher Education (decision no.\ 1266/MOB/IV/2015/0). Work at the University of Chicago is supported in part by U.S. Department
of Energy grant number DE-FG02-13ER41958.
Work at ANL is supported in part by the U.S. Department of Energy under 
Contract No. DE-AC02-06CH11357.


\end{document}